\documentclass[letterpaper]{article} 
\usepackage[]{aaai2026}  
\usepackage{times}  
\usepackage{helvet}  
\usepackage{courier}  
\usepackage[hyphens]{url}  
\usepackage{graphicx} 
\usepackage{natbib}  
\usepackage{caption} 
\usepackage{algorithm}
\usepackage{algorithmic}

\usepackage{newfloat}
\usepackage{listings}
\DeclareCaptionStyle{ruled}{labelfont=normalfont,labelsep=colon,strut=off} 
\floatstyle{ruled}
\newfloat{listing}{tb}{lst}{}
\floatname{listing}{Listing}
\usepackage{amsmath, amssymb, mathtools, amsthm, bm}
\usepackage{graphicx}
\usepackage{subfigure}
\usepackage{verbatim}
\usepackage{soul}
\usepackage[]{footmisc}
\usepackage{enumitem}
\usepackage{tikz}
\usepackage[]{footmisc}

\usetikzlibrary{shapes,arrows}
\usepackage{xpatch}
\usepackage{hhline}
\usepackage{multirow}
\usepackage{algorithm,algorithmic}
\usepackage{cite}

\def\E{\mathbb E }
\newcommand{\real}{\mathbb{R}}

\usepackage{cleveref}
\usepackage{tcolorbox}
\usepackage{diagbox}

\usepackage{fancyhdr}
\title{RUL-QMoE: Multiple Non-crossing Quantile Mixture-of-Experts for Probabilistic Remaining Useful Life Predictions of Varying Battery Materials}
\author{
    Sel Ly,
    Rufan Yang,
    Ninad Dixit, 
    Hung Dinh Nguyen 
}
\affiliations{

    Nanyang Technological University, Singapore 639798 \\
    \{sel.ly, hunghtd\}@ntu.edu.sg, \{rufan001, ninad001\}@e.ntu.edu.sg
}

\begin{document}

\maketitle

\thispagestyle{fancy} 

\begin{abstract}

Lithium-ion (Li-ion) batteries are the major type of battery used in a variety of everyday applications, including electric vehicles (EVs), mobile devices, and energy storage systems. Predicting the Remaining Useful Life (RUL) of Li-ion batteries is crucial for ensuring their reliability, safety, and cost-effectiveness in battery-powered systems. The materials used for the battery cathodes and their designs play a significant role in determining the degradation rates and RUL, as they lead to distinct electrochemical reactions. Unfortunately, RUL prediction models often overlook the cathode materials and designs to simplify the model-building process, ignoring the effects of these electrochemical reactions. Other reasons are that specifications related to battery materials may not always be readily available, and a battery might consist of a mix of different materials. As a result, the predictive models that are developed often lack generalizability. To tackle these challenges, this paper proposes a novel material-based Mixture-of-Experts (MoE) approach for predicting the RUL of batteries, specifically addressing the complexities associated with heterogeneous battery chemistries. The MoE is integrated into a probabilistic framework, called \textit{Multiple Non-crossing Quantile Mixture-of-Experts for Probabilistic Prediction (RUL-QMoE)}, which accommodates battery operational conditions and enables uncertainty quantification. The RUL-QMoE model integrates specialized expert networks for five battery types: LFP, NCA, NMC, LCO, and NMC-LCO, within a gating mechanism that dynamically assigns relevance based on the battery's input features. Furthermore, by leveraging non-crossing quantile regression, the proposed RUL-QMoE produces coherent and interpretable predictive distributions of the battery's RUL, enabling robust uncertainty quantification in the battery's RUL prediction. Trained on seven real-world datasets, the proposed RUL-QMoE achieves strong predictive performance across all battery types,  with MAE =  65 (cycles), MAPE = 9.59\%, RMSE = 100 (cycles), and $R^2=96.84\%$.  Compared to traditional models like XGBoost, Random Forest, CNN, and LSTM, the proposed RUL-QMoE model consistently delivers lower RMSE and superior probabilistic insights, including survival probabilities and prediction intervals. \textcolor{black}{The model has been integrated into our Battery AI platform in collaboration with Toyota Motor Engineering \& Manufacturing North America, Inc., as part of a broader Battery Foundation Model initiative. This RUL-QMoE model will serve future Toyota EVs' users and battery system designers.}
\end{abstract}

%

\pagestyle{plain} 
\section{Introduction}

Lithium-ion batteries underpin a wide spectrum of modern technologies, from electric vehicles and renewable energy storage to portable electronics \cite{xie2024dual}. However, their performance inevitably declines due to degradation mechanisms such as capacity fade, abnormal heat generation, and internal resistance growth \cite{yang2025temperature}. These phenomena not only reduce operational efficiency but also raise safety concerns, particularly as batteries approach end-of-life \cite{xie2022health}. Accurate prediction of Remaining Useful Life (RUL) is therefore essential for ensuring safety, reliability, and economic efficiency. By forecasting the time or cycles before a battery’s performance falls below acceptable thresholds, RUL prediction supports proactive maintenance scheduling, mitigates the risk of sudden failures, and optimizes end-of-life management for both economic and environmental benefits.

Traditional approaches to RUL prediction fall into three categories: physics-based models, data-driven models, and hybrid models \cite{hasib2021comprehensive}. Physics-based methods capture electrochemical and thermal processes in detail but are hindered by computational complexity and high sensitivity to parameter uncertainty \cite{wang2023research}. Data-driven methods, powered by machine learning (ML), bypass explicit modelling by learning patterns directly from historical and real-time data. They offer flexibility and scalability but demand large, high-quality datasets and often fail to generalize across battery chemistries due to dataset heterogeneity and feature engineering limitations \cite{zhangbatteryml}. Hybrid approaches seek to merge the interpretability of physics-based models with the flexibility of ML, yet integration challenges and inadequate uncertainty quantification remain unresolved. What is more, existing RUL prediction methods failed to address two problems: 1. Material-specific degradation: Existing models are typically tailored to a single type of Lithium-ion battery. However, different types of Lithium-ion batteries have different degradation and operational features. For instance, the LFP batteries exhibit flat voltage plateaus, while NMC shows steep capacity drops. 2. Uncertainty propagation: Existing methods provide point estimates of the battery RUL without confidence intervals, which are inaccurate since the factors to RUL are uncertain during the operational life of the targeted Li-ion battery.

Artificial intelligence (AI) is now reshaping the RUL prediction landscape. Advances in deep learning, probabilistic modelling, and mixture-of-experts (MoE) architectures enable AI systems to learn degradation patterns from diverse datasets, adapt to varying chemistries and usage profiles, and provide probabilistic forecasts that account for uncertainty. Such capabilities are critical for real-world applications, where battery chemistries, such as Lithium Iron Phosphate (LFP), Nickel Cobalt Aluminum (NCA), Nickel Manganese Cobalt (NMC), and Lithium Cobalt Oxide (LCO) - exhibit distinct degradation signatures. Moreover, AI-based frameworks can be designed for continuous learning, improving prediction accuracy as new data becomes available.

In this study, we propose RUL-QMoE, a Multiple Non-crossing Quantile Mixture-of-Experts framework for probabilistic battery RUL prediction. Each expert model is specialized for a specific battery chemistry, while a gating network dynamically assigns relevance based on input features. By predicting multiple quantiles of the RUL distribution, RUL-QMoE offers a comprehensive view of potential outcomes rather than single-point estimates, enabling robust uncertainty quantification. To ensure coherent probabilistic outputs, the model enforces monotonicity across quantiles through a non-crossing constraint. Leveraging seven diverse real-world datasets, including CALCE \cite{xing2013ensemble,he2011prognostics}, MATR \cite{severson2019data,hong2020towards}, HUST \cite{ma2022real}, HNEI \cite{devie2018intrinsic}, RWTH \cite{li2021one}, SNL \cite{preger2020degradation} and UR-PUR \cite{juarez2020degradation,juarez2021degradation} - our framework demonstrates high adaptability, scalability, and accuracy across chemistries and operational profiles.

By integrating AI-based tools, probabilistic forecasting, and chemistry-specific expertise, RUL-QMoE advances the state-of-the-art in battery health management. \textcolor{black}{The RUL-QMoE model is currently being deployed as part of the Battery Foundation Model initiative in collaboration with Toyota Motor Engineering \& Manufacturing North America, Inc., where it powers Battery AI agents for predictive maintenance in EVs}. Also, it will lay the foundation for intelligent, data-adaptive, and risk-aware predictive maintenance systems that can scale with evolving battery technologies, accelerating the transition toward safer, more reliable, and more sustainable energy storage solutions. 

\section{Related Work}
Recent works in data-driven RUL prediction for lithium-ion batteries leverage historical or real-time datasets to learn degradation behaviours without relying on explicit physical models. For instance, \cite{catelani2021remaining} proposed a hybrid approach that combines a physics-based single exponential degradation model with a deep echo state network, optimized via genetic algorithms, and validated using NASA's lithium-ion datasets. Similarly, \cite{yang2021machine} developed a hybrid 3D-2D convolutional neural network with feature and multi-scale cycle attention mechanisms to predict battery cycle life and RUL, utilizing voltage, current, and temperature curves from charge cycles, and achieving test errors of 1.1\% and 3.6\% under diverse operating conditions. While these models demonstrate the potential of deep learning to enhance prediction accuracy, their architectures are often monolithic, limiting adaptability to heterogeneous datasets and diverse degradation patterns.

To address these limitations, recent advancements have explored Mixture-of-Experts (MoE) architectures, which are particularly effective for battery forecasting tasks due to their ability to partition the learning process into specialized sub-models, each capturing distinct aspects of degradation. In \cite{lei2025multi}, the authors introduced \textit{MSPMLP}, a multi-scale patch-based MLP combined with sparsely-gated MoE to forecast battery capacity, effectively modelling both long-term degradation and short-term recovery, and achieving a 41.8\% improvement in MAPE over prior methods. Likewise, \cite{chen2024attmoe} proposed \textit{AttMoE}, integrating transformer-based attention with MoE to enhance faded capacity prediction under noisy sensor conditions, yielding 10–20\% improvements in relative error. Furthermore, \cite{zhao2024mlp} presented \textit{MMMe}, a hybrid model combining bidirectional GRUs, attention modules, and MLP-Mixer blocks with a 32-expert MoE layer, excelling in modelling both gradual degradation and sudden capacity drops with a RMSE of 0.0515 on the NASA dataset. These studies collectively underscore the versatility of MoE frameworks in handling heterogeneous data and capturing complex degradation behaviours. However, most existing MoE models focus on single-point predictions or are tailored to specific chemistries. In contrast, our proposed RUL-QMoE advances the field by delivering probabilistic RUL predictions across diverse battery types, ensuring non-crossing quantile outputs and robust uncertainty quantification.

\section{Problem Formulation}
In this section, we briefly introduce the task of constructing a predictive model for the conditional quantile of a response variable $Y$, given an input $\bm{X}=\bm{x}$, denoted as $Q_{Y|\bm{X}=\bm{x}}(\tau)$. Here, $Y$ represents the remaining useful life (RUL) of a battery, defined as the number of cycles the battery can continue operating before its capacity drops below a specified threshold (e.g., 80\% of its nominal capacity). The input vector $\bm{X}$ comprises features that influence the battery’s RUL.

\begin{figure}[tb]
    \centering
    \includegraphics[scale = 0.5]{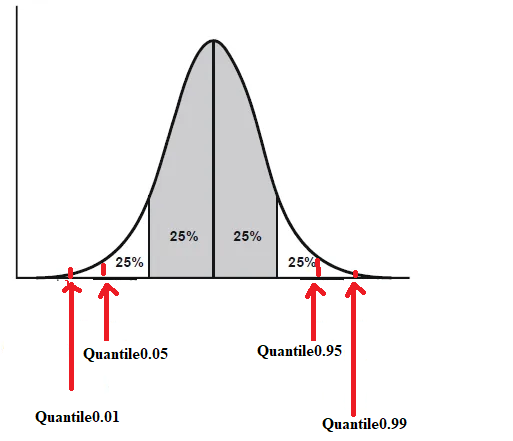}
    \caption{Quantile definition: $Q_Y(\tau)=F^{-1}_Y(\tau), \tau \in (0,1)$.}
    \label{fig:quantile}
\end{figure}

A basic approach to this problem is the quantile regression model proposed in \cite{koenker1978regression}. The quantile $Q_Y(\tau)$ at level $\tau \in (0,1)$
 is defined as the generalized inverse of the cumulative distribution function (CDF) $F_Y(y):=\mathbb{P}(Y \le y)$, see Figure \ref{fig:quantile}:
\begin{align}
    Q_{Y}(\tau) :=F_Y^{-1}(\tau) = \inf\{y \in \real: \mathbb{P}\big(Y\le y \big) \ge \tau \}.
\end{align}

 Accordingly, the conditional quantile of $Y$ given $\bm{X}=\bm{x}$
 is expressed as:
\begin{align}
    Q_{Y|\bm{X}=\bm{x}}(\tau) = \inf\{y \in \real: \mathbb{P}\big(Y\le y \big| \bm{X}=\bm{x}\big) \ge \tau \}.
\end{align}
When $\tau=0.5$, this corresponds to the conditional median of $Y$. By selecting multiple quantile levels, such as $ \tau \in \{0.05,0.1,0.2,…,0.9,0.95\}$, we can generate a range of quantile estimates for $Y$, thereby enabling the construction of $F_{Y|\bm{X}=\bm{x}}(y)$ for probabilistic forecasting.

In this work, we aim to develop a general Mixture-of-Experts model for estimating the conditional distribution of a battery’s RUL, applicable across various chemistries such as LFP, NCA, NMC, LCO, and mixed NMC-LCO types. This is significantly important because each battery chemistry exhibits distinct RUL distributions, and conventional machine learning models typically capture only a single distribution per battery type.

Moreover, when trained on multiple quantile levels, standard models often violate the non-decreasing property of quantiles, leading to the quantile crossing problem. To address this, we design our MoE framework to ensure monotonicity across quantile outputs. Specifically, for any two quantile levels $\tau_1 < \tau_2$, the model guarantees that the predicted quantiles satisfy 
$Q_{Y|\bm{X}=\bm{x}}(\tau_1) < Q_{Y|\bm{X}=\bm{x}}(\tau_2)$. This constraint preserves the natural ordering of quantiles and contributes to more accurate and coherent predictive distributions.

\section{Proposed Model}
\begin{figure}[tb]
    \centering
    \includegraphics[scale=0.42]{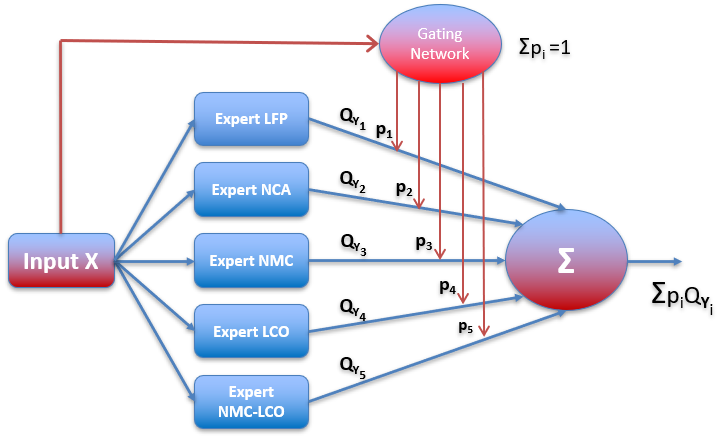}
    \caption{Proposed QMoE architecture for varying types of batteries, where $Q_{Y_i}=Q_{Y_i|\bm{X}=\bm{x}}$ for short.}
    \label{fig:MoE}
\end{figure}

As aforementioned, we propose a novel MoE model that integrates five specialized Expert networks, each corresponding to a specific battery chemistry type: LFP, NCA, NMC, LC, and mixed NMC-LCO. The second core of the MoE model is a Gating network, which learns to assign dynamic, input-dependent weights to each expert, see Figure \ref{fig:MoE} for an overview.

Unlike traditional MoE models that provide single-point predictions using conditional mean values, we propose a multi-quantile MoE approach. This proposed model can output several quantiles of the battery's RUL, denoted as $ Q_{Y|\bm{X}=\bm{x}} = \big(Q_{Y|\bm{X}=\bm{x}}(\tau_1), \ldots, Q_{Y|\bm{X}=\bm{x}}(\tau_K)\big)^T$, where $K$ is the number of quantile levels. Thereby, it enables us to construct the entire conditional distribution of the battery's RUL for a given input $\bm{X}=\bm{x}$. Hence, the proposed model offers a comprehensive uncertainty quantification for RUL predictions. For your reference, we detail the architectures of the proposed Expert and Gating Networks, along with the methods used to construct the predictive battery's RUL distributions in the appendices.
\section{Experiments}

\subsection{Datasets}

To develop and evaluate the proposed RUL-QMoE model, we integrated seven real-world datasets, as summarized in Table \ref{tab:datasets}. These datasets were utilized for feature extraction, model training, and performance evaluation. For a more detailed description of each dataset, we refer readers to the Appendix.

\begin{table*}[tb]
    \centering
 \resizebox{1.9\columnwidth}{!}{   
 \begin{tabular}{l|l|cccc}
        \textbf{Dataset}	& \textbf{Electrode Chemistry} &	\textbf{Rated Capacity (Ah)}	& \textbf{Voltage Range (V)}	& \textbf{RUL Distr. (Cycles)} &	\textbf{Cell Count} \\ \hline
CALCE &	LCO/graphite	& 1.1	& 2.7 - 4.2	& $566 \pm 106$	& 13 \\
MATR	& LFP/graphite	& 1.1	& 2.0 - 3.6 &	$823 \pm 368$	& 180 \\ 
HUST	& LFP/graphite &	1.1	& 2.0 - 3.6	& $1899 \pm 389$	& 77 \\
HNEI	& NMC LCO/graphite	& 2.8	& 3.0 - 4.3 &	$248 \pm 15$	& 14 \\
RWTH	& NMC/carbon &	1.11 &	3.5 - 3.9 &	$658 \pm 64$ &	48 \\
SNL	& NCA, NMC, LFP/graphite	& 1.1	& 2.0 - 3.6	& $1256 \pm 1321$	& 61 \\
UL-PUR &	NCA/graphite	& 3.4	& 2.7 - 4.2	& $209 \pm 50$ &	10 \\ \hline
    \end{tabular}
}
\caption{Battery Datasets \cite{zhangbatteryml}.}
    \label{tab:datasets}
\end{table*}

\subsection{Feature Engineering}

The battery datasets described above are highly heterogeneous, with numerous factors influencing lifespan, such as chemistry, capacity, voltage range, and cycling conditions. While prior research has proposed various feature extraction strategies, these methods are often tailored to specific datasets and lack generalizability across chemistries and operational profiles \cite{zhangbatteryml}.

In this work, we unify seven publicly available datasets to construct a comprehensive mixed dataset. For raw data preprocessing, we adopt the standardized pipeline provided by \cite{zhangbatteryml}. To construct the input feature vector $\bm{X}$ for model training, we include: one-hot encoded cathode material (covering five types), nominal capacity, minimum and maximum voltage, charge and discharge C-rates, and a sequence of discharge voltage-capacity curves from cycle 10 to cycle 100, denoted as $V_{d,t}(Q_{d,t})$ for $t = 10, \ldots, 100$.

\begin{figure}[tb]
    \centering
    \includegraphics[scale=0.38]{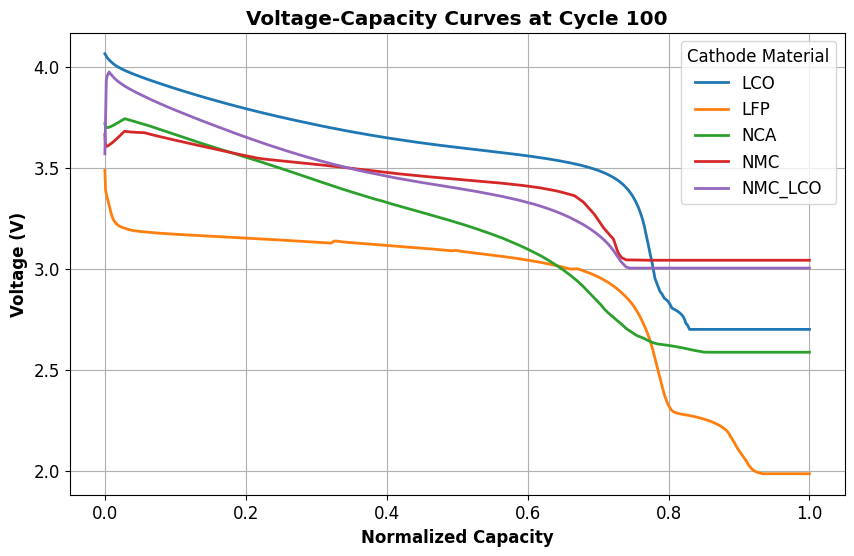}
    \caption{ Means of discharge voltage-capacity curves at cycle 100 for varying types of batteries.}
    \label{fig:X_mean}
\end{figure}

To ensure consistency across batteries, we interpolate each discharge curve to 1000 fixed capacity points. This results in a feature vector $\bm{X}$ of dimension 1010. Figure~\ref{fig:X_mean} illustrates the mean interpolated voltage-capacity curves for each cathode material, highlighting distinct degradation signatures across chemistries.

\subsection{Implementation of Model Training}

We train the proposed RUL-QMoE model in two stages to ensure both specialization and generalization across battery chemistries.

In \textbf{Stage 1}, we independently train five Expert Networks, each corresponding to a specific cathode material type: LFP, NCA, NMC, LCO, and NMC-LCO. Each expert is optimized using the Adam optimizer with a learning rate of 0.001, a batch size of 256, and 256 hidden units. The dataset is split into 70\% for training and 30\% for testing, with a fixed random seed of 42 to ensure reproducibility. To prevent overfitting, we employ an Early Stopping strategy with a patience of 20 epochs, halting training when validation performance ceases to improve.

In \textbf{Stage 2}, the pre-trained experts are used as warm-start components, and the Gating Network is trained jointly with fine-tuning of the expert parameters. This stage enables the model to learn optimal expert combinations for each input while refining the experts’ predictions. Note that we conducted all training on a free T4 GPU-enabled environment using Google Colab to accelerate computation.

\subsection{Evaluation Metrics}
To evaluate the performance of the trained RUL-QMoE model, we use the conditional median prediction, denoted as $Q_{Y|\bm{X}=\bm{x}}(0.5)$, as the representative single-point estimate of the battery's RUL. We assess prediction accuracy using standard regression metrics, including Mean Absolute Error (MAE), Mean Absolute Percentage Error (MAPE), Root Mean Square Error (RMSE), and the Coefficient of Determination ($R^2$). These metrics provide a comprehensive view of both absolute and relative error, as well as the model’s goodness-of-fit across diverse battery chemistries and datasets.


\subsection{Model Performance}



The proposed RUL-QMoE model demonstrates strong predictive performance across a range of battery chemistries, with particularly high accuracy observed for NMC-LCO and LFP types as shown in Table \ref{tab:bat_type}. The lowest MAE is achieved on the mixed NMC-LCO battery type (MAE = 10, MAPE = 5.2\%), indicating highly precise single-point predictions. Moreover, the model achieves its highest coefficient of determination ($R^2$ = 98.34\%) on the ALL dataset (containing all battery's cathode materials), highlighting its ability to generalize effectively across diverse battery types. In contrast, the model's performance on LCO batteries is comparatively weaker, with the lowest $R^2$ value of 33.22\%. This discrepancy could be due to a small sample size of the LCO dataset, and thus, insufficient feature representation for this chemistry. However, the model maintains robust performance across most battery types, making it a promising tool for RUL prediction in heterogeneous battery applications.  
\begin{table}[tb]
    \centering
    \begin{tabular}{l|cccc}
        \textbf{Battery Type} & 	\textbf{MAE}	& \textbf{MAPE}	&  \textbf{RMSE} &	$\bm{R}^2$ \\ \hline
LFP	& 62	& 7.01\%	& 116	& 97.85\% \\
NCA &	36	& 13.80\%	& 57	& 92.26\% \\
NMC	& 47 &	8.83\%	& 61	& 83.83\% \\
LCO	& 51	& 8.96\%	& 82	& 33.22\% \\
NMC-LCO &	10 &	5.20\% &	12	& 83.43\% \\ 
\textbf{ALL}	& 55	& 7.82\%	& 98	& 98.34\% \\
\hline
    \end{tabular}
    \caption{Performance of the proposed RUL-QMoE model for varying types of batteries.}
    \label{tab:bat_type}
\end{table}

\begin{figure*}[htb]
    \centering
    \includegraphics[scale=0.4]{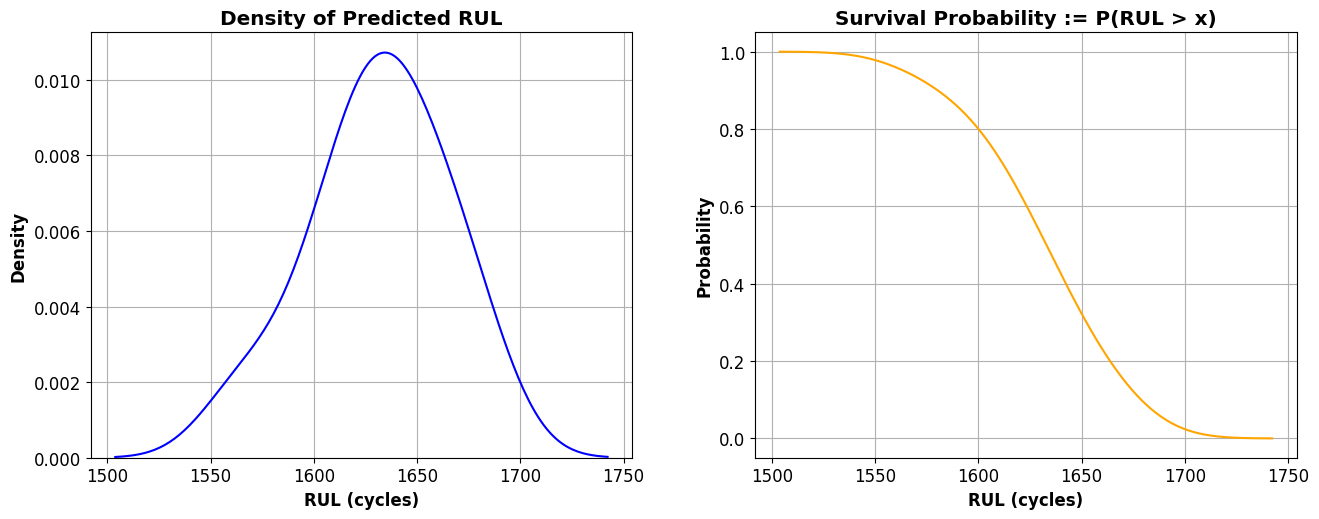}
    \caption{ Probabilistic battery's RUL Predictions.}
    \label{fig:prob_pred}
\end{figure*}

As mentioned, one notable advantage of our proposed model is its ability to provide uncertainty quantification, such as prediction intervals (PI) and survival probabilities. For instance, Figure \ref{fig:prob_pred} illustrates the probabilistic prediction capabilities for a specific battery, with a predicted median RUL of 1632 cycles, and a 90\% PI ranging from 1565 to 1681 cycles, which covers the ground truth RUL of 1599 cycles. Furthermore, we can also perform survival analysis at a specific cycle, for instance, at cycle 100.  If the model estimates a survival probability of $\mathbb{P}(\text{RUL} > 1600) = 0.8$, this indicates that the battery has an 80\% chance of operating for at least an additional 1600 cycles beyond cycle 100.


\subsection{Comparisons with Other Models}

\begin{figure*}[h]
    \centering
    \includegraphics[scale=0.4]{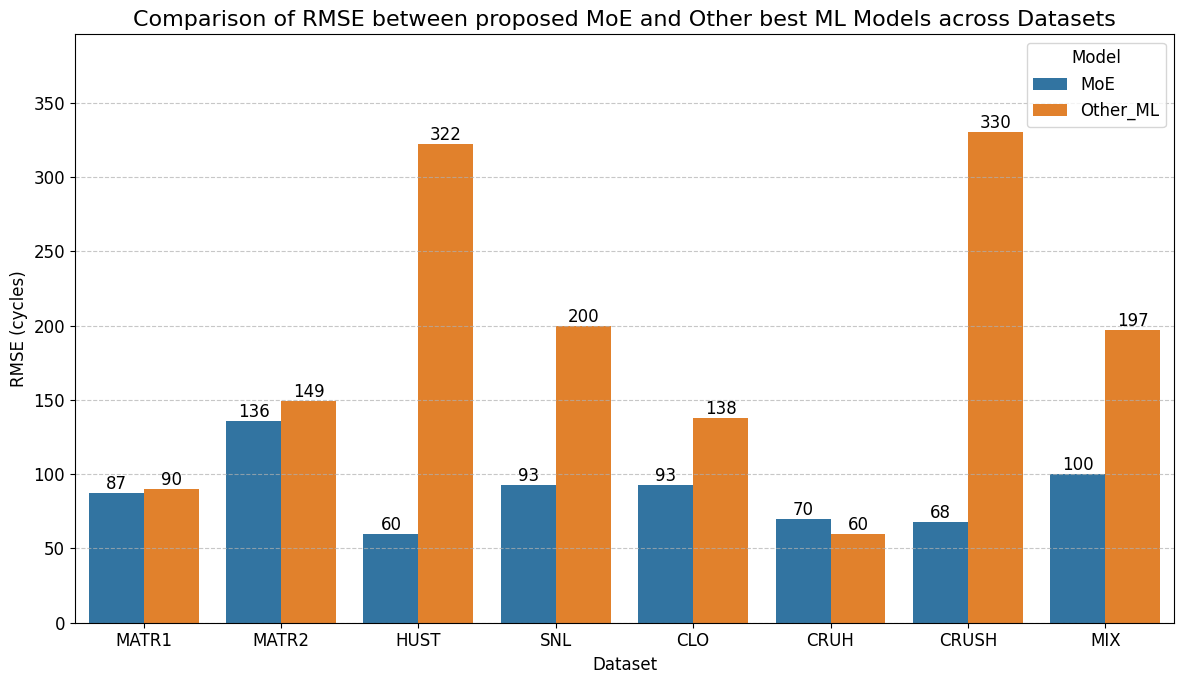}
    \caption{Performance comparison between the proposed RUL-QMoE and other best ML models in \cite{zhangbatteryml} for RUL predictions across many testing datasets.} 
    \label{fig:rmse_comparison}
\end{figure*}

To benchmark the performance of the proposed RUL-QMoE model, we shall test and compare our proposed model against several leading ML models in \cite{zhangbatteryml} across eight battery datasets using cycling data at cycle 100 only.
As can be seen from Table \ref{tab:bat_data}, the best performance is observed on the HUST dataset, which achieves the lowest MAE of 47 (cycles) and MAPE of 2.56\%. Similarly, the CRUSH dataset shows excellent results with a low MAE of 49 (cycles). In contrast, the MATR2 dataset presents the greatest challenge, with the highest MAE of 98 (cycles), RMSE of 136 (cycles), and the lowest $R^2$ of 80.13\%. This suggests that the dataset may contain greater variability or noise, making it more difficult for the model to learn reliable degradation patterns. For the MIX dataset, which combines all battery datasets, the model maintains strong performance, with MAE = 65 (cycles), RMSE = 100 (cycles), and $R^2$ = 96.84\%, validating its scalability and robustness across chemistries and operational conditions. These results confirm that the proposed RUL-QMoE is capable of delivering accurate RUL predictions from early-cycle data, making it a valuable tool for early-stage battery health assessment and predictive maintenance.
\begin{table}[t]
    \centering
    \begin{tabular}{l|cccc}
        \textbf{Battery Data} & 	\textbf{MAE}	& \textbf{MAPE}	&  \textbf{RMSE} &	$\bm{R}^2$ \\ \hline
MATR1	& 53	& 8.05\%	& 87	& 94.93\% \\
MATR2	& 98 &	10.19\% &	136	& 80.13\% \\
HUST	& 47 &	2.56\% &	60 &	97.93\%\\
SNL	& 67	& 9.94\% &	93	& 99.44\% \\
CLO	& 65 &	8.92\% &	93	& 92.12\% \\
CRUH	& 53 &	12.64\% &	70	& 84.55\% \\
CRUSH	& 49 &	10.61\%	& 68 &	99.55\% \\
MIX	& 65	& 9.59\% &	100 &	96.84\% \\ \hline
    \end{tabular}
\caption{Performance of the proposed RUL-QMoE model
for several datasets of batteries using data at cycle 100 only.} \label{tab:bat_data}
\end{table}

As shown in Figure~\ref{fig:rmse_comparison}, the proposed RUL-QMoE consistently achieves lower RMSE values, compared to best ML models implemented in \cite{zhangbatteryml}. Specifically, on the HUST dataset, the proposed RUL-QMoE delivers a dramatic improvement with an RMSE of 60 cycles, compared to 322 cycles from the best ML model, namely, \textit{''Discharge model''} in \cite{zhangbatteryml}. For CRUSH, the model achieves an RMSE of 100 cycles, significantly outperforming the XGBoost model in \cite{zhangbatteryml} which yields an RMSE of 330 cycles. On CLO, RUL-QMoE reduces RMSE to 70 cycles versus 138 cycles from competing \textit{``Full'' model}. For the MIX dataset, which combine all battery datasets, the RUL-QMoE model achieves an RMSE of 100 cycles - nearly halving the error compared to 197 cycles from the Random Forest model. A similar trend is observed on the SNL dataset, where RUL-QMoE reduces the RMSE to 93 cycles, significantly outperforming the Principal Component Regression (PCR) model, which records an RMSE of 200 cycles. Also, moderate improvements are also observed on MATR1, and MATR2, with RMSE reductions of 3 and 15 cycles, respectively. However, on the CRUH dataset, Partial Least Squares Regression (PLSR) slightly outperforms the proposed RUL-QMoE model, achieving a lower RMSE of 60 cycles compared to 70 cycles. 


These results underscore the effectiveness of the RUL-QMoE framework in delivering accurate and generalizable RUL predictions across diverse battery chemistries and operational conditions. Besides, its consistent performance advantage over existing ML models validates the strength of its probabilistic, expert-driven architecture.

\section{Path to Deployment}
The proposed RUL-QMoE framework presents a significant advancement in probabilistic battery health and prognostics, with clear pathways for real-world deployment across applications reliant on lithium-ion batteries. Its architecture, performance, and uncertainty quantification capabilities align well with the operational needs in existing Li-ion battery EV systems. We outline the key deployment considerations and strategies below:
\begin{itemize}
    \item \textbf{Seamless Integration with Battery Management Systems (BMS)}\\
    For EVs and portable devices which has limited computational resources for BMS, RUL-QMoE can be compressed and deployed on edge devices and use early-cycle data from the BMS to evaluate the model and minimize the computational latency, enabling real-time and accurate RUL updates during operation. What is more, in large-scale storage systems such as grid storage, lightweight gating networks can be trained locally to assign experts, while computationally-intensive tasks like quantile predictions are offloaded to the cloud.
    \item \textbf{Uncertainty-Aware Maintenance and Operations}\\
    The probabilistic outputs of RUL-QMoE, including survival probabilities, non-crossing prediction intervals, and full conditional distributions - transform uncertainty quantification into actionable intelligence for risk-informed maintenance. By dynamically estimating the survival probability of a battery at any cycle, the model enables adaptive replacement triggers based on user-defined risk thresholds, further supports conservative maintenance scheduling, ensuring replacements occur before the lower prediction interval bound to prevent failures while optimizing spare-parts logistics. Integration with industrial IoT platforms automates alerting and work-order generation in Computerized Maintenance Management Systems when survival probabilities reach operational limits or observed degradation deviates beyond prediction intervals.
    \item \textbf{Adaptability to Emerging Chemistries and Data Streams} \\
    The MoE architecture supports seamless addition of new expert networks for emerging battery chemistries. Incremental fine-tuning with federated learning can preserve data privacy across manufacturers. Also, the pretrained experts can accelerate adaptation to new datasets with limited samples, reducing data acquisition costs.
\end{itemize}
RUL-QMoE bridges the gap between academic research and industrial deployment. Its chemistry-specific expertise, probabilistic outputs, and adaptability position it as a cornerstone for next-generation battery health management. Pilot deployments in collaboration with automotive and energy partners will be the critical next step toward commercialization.

\section{Conclusion and Future Work}

We presented RUL-QMoE, a novel probabilistic Mixture-of-Experts framework for estimating the Remaining Useful Life (RUL) of lithium-ion batteries across diverse chemistries and operational conditions. Our approach integrates specialized expert networks aligned with distinct cathode materials and enforces non-crossing quantile constraints to produce coherent, interpretable probabilistic forecasts. A dynamic gating mechanism enables adaptive relevance assignment, allowing the model to generalize effectively across heterogeneous datasets. Comprehensive evaluations on seven publicly available battery datasets demonstrate that RUL-QMoE consistently surpasses state-of-the-art baselines in RMSE, MAE, and $R^2$, with notable gains on mixed-chemistry and early-cycle data. Beyond predictive accuracy, the model supports survival analysis and prediction interval estimation - critical capabilities for risk-aware battery management and maintenance planning.

\textcolor{black}{RUL-QMoE offers a scalable, data-efficient foundation for battery health prognostics, with strong potential for real-world deployment in electric vehicles, grid storage systems, and portable electronics. It serves as a core component of the Battery Foundation Model currently under development in collaboration with Toyota Motor Engineering \& Manufacturing North America, Inc., and has been integrated into our Battery AI platform during Stage 1 of the project.}

\textcolor{black}{In Stage 2, we aim to extend the framework to monitor additional battery states, including State-of-Health (SoH) and State-of-Charge (SoC), and to support downstream applications such as estimating EV driving range under varying conditions (e.g., weather, traffic, battery specifications, vehicle brand, RUL, SoH, SoC, etc.). These enhancements will further position RUL-QMoE as a key enabler of intelligent, context-aware energy systems, advancing the deployment of AI in sustainable mobility and energy infrastructure.}

\section*{Acknowledgments}
This research is supported by the Agency for Science, Technology and Research (A*STAR), Singapore under its project M23M6c0114, and Toyota Motor Engineering \& Manufacturing North America, Inc.

\section*{Appendix: RUL-MoE Architecture}
\subsection*{Proposed Multiple Non-crossing Quantile Expert Networks}

\begin{figure}[tb]
    \centering
    \includegraphics[scale=0.4]{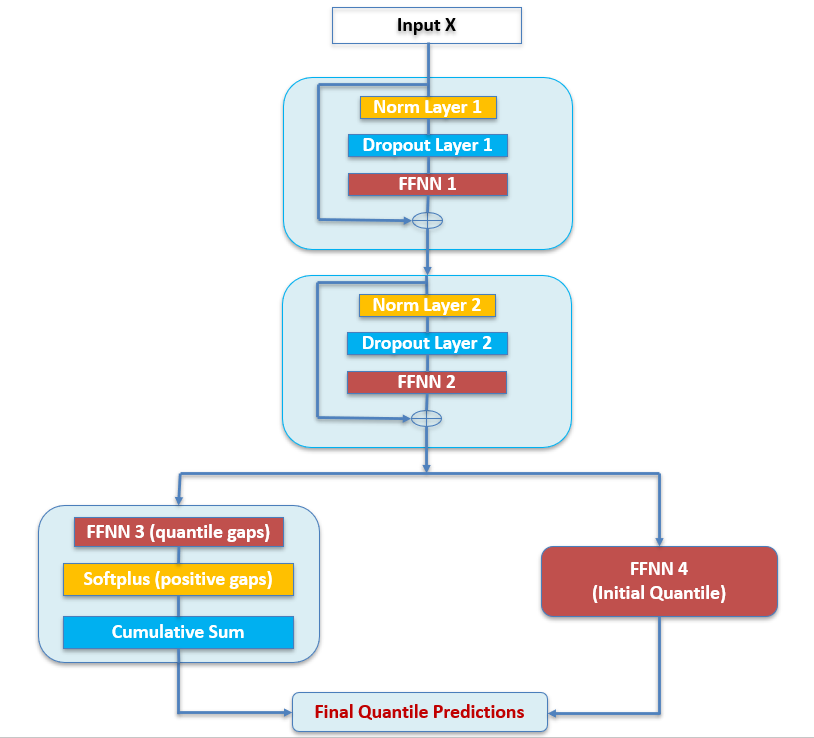}
    \caption{Expert network architecture for each battery type.}
    \label{fig:expert}
\end{figure}

In this work, we design each Expert network tailored for each battery’s cathode material using multiple non-crossing quantile neural network regression for battery's RUL prediction, see Figure \ref{fig:expert}. Specifically, the architecture of each expert neural network begins with an input vector $\bm{X}=\bm{x}$, which is processed through two sequential blocks. Each block consists of:

\begin{itemize}
    \item \textbf{Normalization Layer}: It standardises input features to zero mean and unit variance. This helps us to stabilize training by reducing internal covariate shift and ensure consistent input distributions across layers:
    \begin{align}
        \bm{x} =\frac{\bm{x} -\E[\bm{x}]}{\sqrt{\text{Var}[\bm{x}]+\varepsilon}} \gamma +\beta, \nonumber 
    \end{align}
    where $\gamma$ and $\beta$ are learnable parameters of the normalization layer.
    \item \textbf{Dropout Layer (optional)}: It randomly deactivates a fraction of neurons (dropout rate) during training to prevent overfitting. This encourages the network to learn robust and redundant representations, improving generalization. Here, we can set the dropout rate to zero to disable this layer. 
    \item \textbf{Feedforward Neural Network (FFNN) with Skip Connection}: A fully connected layer with non-linear activation functions that extracts and transforms features relevant to battery degradation. Also, each FFNN block includes a residual connection that adds the input of the block to its output. This skip connection facilitates gradient flow, mitigates vanishing gradient issues, and improves training stability:
    \begin{align}
\bm{h}_i^{(1)}&:=\bm{x} + \text{ReLU}\big(\bm{W}_i^{(1)} \bm{x} + \bm{B}_i^{(1)}  \big), \, i =1,\ldots5,  \nonumber \\
\bm{h}_i^{(2)}&:=\bm{h}_i^{(1)}+\text{ReLU}\big(\bm{W}_i^{(2)} \bm{h}_i^{(1)} + \bm{B}_i^{(2)}  \big), \, i =1,\ldots, 5, \nonumber 
\end{align}
where the well-known ReLU activation function is defined by $\text{ReLU}(x):= \max(0,x)$ for all $x \in \mathbb{R}$.
\end{itemize}

After these two blocks, the network splits into two branches:

\begin{enumerate}
    \item \textbf{Quantile Gap Branch}: This branch uses an additional FFNN 3 to predict strictly positive quantile gaps. A \texttt{Softplus} activation function ensures all outputs are positive: 
    \begin{align}
    \bm{h}_i^{(3)} &:= \text{Softplus}\big(\bm{W}_i^{(3)} \bm{h}_i^{(3)} + \bm{B}_i^{(3)}\big). \nonumber
\end{align}
These gaps are then cumulatively summed to construct a sequence of increasing quantile values, enforcing non-crossing behaviour, see \eqref{sec:K-outputs}.
    \item \textbf{Initial Quantile Branch}: A separate FFNN 4 predicts the base quantile value, which serves as the starting point for the cumulative quantile construction:
    \begin{align}
    h_i^{(4)} &:= \bm{W}_i^{(4)} \bm{h}_i^{(4)} + B_i^{(4)}. \nonumber
    \end{align}
\end{enumerate}
.

To enforce non-crossing behaviour across multiple quantiles, say, $0<\tau_1 < \tau_2 < \ldots < \tau_K <1$, we cumulatively sum the predicted quantile gaps (FFNN 3) along the quantile axis and add them to the initial quantile estimate (FFNN 4):
\begin{align}
    Q_{Y_i|\bm{X}=\bm{x}}(\tau_k) =    \sum \limits_{1 \le j \le k} h_{i,j}^{(3)} + h_i^{(4)}, \, k=1,\ldots, K. \label{sec:K-outputs} 
\end{align}
Indeed, for each quantile level $\tau_k \in (0,1)$, we have that
\begin{align}
    Q_{Y_i|\bm{X}=\bm{x}}(\tau_1) &=   h_{i,1}^{(3)} + h_i^{(4)}, \label{sec:tau1}\\
    Q_{Y_i|\bm{X}=\bm{x}}(\tau_2) &=   \Big(h_{i,1}^{(3)}+h_{i,2}^{(3)}\Big) + h_i^{(4)}, \label{sec:tau2}\\
    \ldots & \ldots \nonumber \\
Q_{Y_i|\bm{X}=\bm{x}}(\tau_K) &=  \Big(h_{i,1}^{(3)}+\ldots+h_{i,K}^{(3)}\Big)   + h_i^{(4)}.  \label{sec:tauK}
\end{align}
Therefore, we obtain final quantile predictions for each expert network $i$, for $i=1,2,3,4,5$:
\begin{align}
    Q_{Y_i|\bm{X}=\bm{x}}(\tau_1) <   Q_{Y_i|\bm{X}=\bm{x}}(\tau_2) < \ldots <  Q_{Y_i|\bm{X}=\bm{x}}(\tau_K). 
\end{align}
This proves that the output quantiles of battery’s RUL are strictly increasing and consistent,(i.e. non-crossing), making the model suitable for stable and interpretable quantile estimation across multiple levels. For example, we take $K=11$, e.g. eleven quantile levels of $\{0.05, 0.1, 0.2,\ldots, 0.9, 0.95\}$. 

\subsection*{Proposed Gating Network}

\begin{figure}[tb]
    \centering
    \includegraphics[scale=0.5]{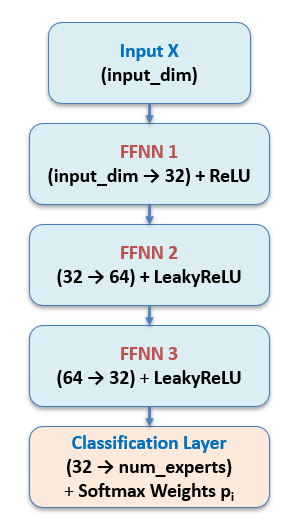}
    \caption{Gating network architecture for proposed MoE.}
    \label{fig:gating}
\end{figure}

In this paper, we propose a Gating network consisting of four linear layers with intermediate LeakyReLU activations: $\text{LeakyReLU}(x)=\max(0,x)+\theta\min(0,x)$, with $\theta$ is called the negative slope. Here, our proposed gating network reduces the input features $\bm{X}$ to a final layer that outputs a softmax probability distribution over the five experts, see Figure \ref{fig:gating}. These weights indicate the relevance of each expert for a given input. 

During the forward pass, each expert processes the same input independently, and their outputs are combined using the learned softmax weights $p_i, \,i=1,2,3,4,5$ from the gating network via a weighted sum: 
\begin{align}
    Q_{Y|\bm{X}=\bm{x}}(\tau_k) =\sum \limits_{i=1}^5 p_i Q_{Y_i|\bm{X}=\bm{x}}(\tau_k), \, k=1,2,\ldots, K. \label{eq:MoE_output1}
\end{align}
This allows the proposed RUL-QMoE model to selectively leverage knowledge from the most appropriate expert(s), enabling more accurate and specialized predictions depending on battery type characteristics. 

Unlike conventional MoE using the mean square loss, we minimize a quantile score (QS) for training our proposed RUL-QMoE, defined as:
\begin{align}
    QS\big(Y,Q_{Y|\bm{X}=\bm{x}}\big) = \frac{1}{K}\sum \limits_{k=1}^K \E\big[\rho_{\tau_k}\big(Y-Q_{Y|\bm{X}=\bm{x}}(\tau_k)\big)\big], \label{eq:QS}
\end{align}
where $\rho_\tau(u) := \max\{\tau u, (\tau-1) u\}, \, \tau \in (0,1),$ is called the pinball loss function.

\section*{Appendix: Estimations of Conditional Predictive Distributions}
We note that the proposed RUL-QMoE has a total of $K$ final quantile outputs:
\begin{align}
    Q_{Y|\bm{X}=\bm{x}} = \big(Q_{Y|\bm{X}=\bm{x}}(\tau_1), \ldots, Q_{Y|\bm{X}=\bm{x}}(\tau_K)\big)^T,
\end{align}
where each $Q_{Y|X=x}(\tau_k)$ defined as in \eqref{eq:MoE_output1}, from which we can estimate the conditional probability density function (PDF) of the RUL variable $Y$ given the input $\bm{X}=\bm{x}$, using a kernel-based smoothing method as follows: 
\begin{align}
    \widehat{f}_{Y|\bm{X}=\bm{x}}(y) = \frac{1}{bK}\sum\limits_{k=1}^{K}\varphi\Big(\frac{y-\widehat{Q}_{Y|\bm{X}=\bm{x}}(\tau_k)}{b}\Big), \label{eq:pdf}  
\end{align}
where $\varphi$ denotes the probability density function of the standard Gaussian distribution (used as the kernel) and $b$ is the bandwidth parameter controlling the smoothness of the estimate. Then, the corresponding conditional CDF given the input $\bm{X}=\bm{x}$, is obtained by integrating the estimated conditional PDF
\begin{align}
    \widehat{F}_{Y|\bm{X} =\bm{x}}(y) &:=\mathbb{P}(Y \le y | \bm{X}= \bm{x}) \nonumber \\
    &= \frac{1}{bK}\sum\limits_{k=1}^{K}\int_{-\infty}^y\varphi\Big(\frac{z-\widehat{Q}_{Y|\bm{X}=\bm{x}}(\tau_k)}{b}\Big)dz.  \label{eq:cdf} 
\end{align}
From this, the conditional survival function is naturally defined as:
\begin{align}
    \widehat{S}_{Y|\bm{X} =\bm{x}}(y):=\mathbb{P}(Y > y | \bm{X}= \bm{x}) =1-\widehat{F}_{Y|\bm{X} =\bm{x}}(y). \label{eq:surv}
\end{align}
These estimated conditional distribution functions $\widehat{f}_{Y|\bm{X} =\bm{x}}(y)$, $\widehat{F}_{Y|\bm{X} =\bm{x}}(y)$, and $\widehat{S}_{Y|\bm{X} =\bm{x}}(y)$ enable rich statistical inference for the battery's RUL variable, including the computation of moments, survival probabilities, prediction intervals (PIs), and other descriptive statistics. For example, a $100(1-\alpha)\%$ PI is derived by two quantile predictions: 
\begin{align}
    \big[Q_{Y|\bm{X}=\bm{x}}(\alpha/2),  Q_{Y|\bm{X}=\bm{x}}(1-\alpha/2)\big].
\end{align}
\section*{Appendix: Datasets}
This appendix provides a more comprehensive overview of all datasets used for training and evaluating the proposed RUL-QMoE model, as summarized in Table \ref{tab:datasets}.

\subsection*{CALCE Dataset}
Sourced from the CS2 and CX2 battery series, the CALCE dataset originates from the Center for Advanced Life Cycle Engineering \cite{xing2013ensemble,he2011prognostics}. These prismatic Lithium Cobalt Oxide (LCO) batteries have a rated capacity of 1.1 Ah. The dataset encompasses complete lifecycle data, with each cell charged using a CC-CV protocol - initially at 0.5C until reaching 4.2V, then held at 4.2V until current drops below 0.05A. Discharge terminates at 2.7V. This consistent setup provides rich data for studying LCO cell degradation under standardized conditions.

\subsection*{MATR Dataset}
This dataset consists of two sets of commercial 18650-format Lithium Iron Phosphate (LFP) cells from various suppliers \cite{severson2019data, hong2020towards}. With 180 cells in total, it is the largest publicly available collection featuring full charge-discharge profiles. The batteries were tested using a 48-channel Arbin LBT system housed in a temperature-controlled chamber at 30°C. Each cell has a nominal 1.1 Ah capacity and 3.3 V rating. Though discharged uniformly, different fast-charging techniques were applied until each cell reached 80\% of its original capacity. Data are grouped into MATR1, MATR2, and CLO subsets by batch.

\subsection*{HUST Dataset}
Containing 77 LFP cells identical in specifications to those in the MATR dataset, the HUST dataset follows the same charging routine but varies its discharge processes using multi-stage methods \cite{ma2022real}. Testing occurred at a constant 30°C. It offers insights into how discharge profiles affect battery degradation, complementing the MATR study.

\subsection*{HNEI Dataset}
The HNEI dataset features commercial 18650 cells with a graphite anode and a mixed cathode of Nickel Manganese Cobalt (NMC) and LCO \cite{devie2018intrinsic}. These cells were cycled at a 1.5C rate to full depth of discharge (100\% DoD) for over 1000 cycles at ambient temperature. The dataset sheds light on how these cathode combinations respond to high-rate cycling and long-term wear.

\subsection*{RWTH Dataset}
This dataset includes 48 lithium-ion cells of the Sanyo/Panasonic UR18650E type, featuring carbon anodes and NMC cathodes \cite{li2021one}. Baseline performance was established via Begin-Of-Life (BOL) testing, followed by periodic Aging Reference Parameter Tests (RPT) to monitor capacity fade and performance changes. The consistent aging protocol allows for detailed analysis of NMC degradation.

\subsection*{SNL Dataset}
This diverse set includes 18650 cells built from NCA, NMC, and LFP chemistries \cite{preger2020degradation}. Cells are cycled to 80\% capacity, with the experiments still in progress. The study explores how temperature, depth of discharge, and discharge current influence aging. Each round of testing comprises a capacity check, several cycles under specific variables, then another check - ideal for modelling real-world degradation patterns.

\subsection*{UL-PUR Dataset}
The UL-PUR dataset consists of graphite/NCA-based pouch cells \cite{juarez2020degradation,juarez2021degradation}. These were cycled at 1C between voltages of 2.7V and 4.2V (0–100\% SOC) at room temperature until capacity dropped by 10–20\%. Additionally, modules were tested at C/2 between 13.7V and 21.0V, stopping after 20\% capacity fade. The data is instrumental for evaluating how different cycling and fade thresholds impact NCA-based systems.

\subsection*{CRUH and CRUSH Composite Datasets}
As described in \cite{zhangbatteryml}, the CRUH dataset combines battery data from CALCE, RWTH, UL-PUR, and HNEI. Extending this integration, CRUSH incorporates cells from the SNL collection alongside the aforementioned sources. These aggregated sets offer a broader basis for advanced modelling and analysis.

\subsection*{MIX Dataset} We combined all available cells from every dataset to form the MIX dataset, which represents the most extensive battery degradation dataset with complete cycling information.

Note that the Battery Archive hosts the SNL \cite{preger2020degradation}, UL-PUR \cite{juarez2020degradation}, and HNEI \cite{devie2018intrinsic} datasets. These datasets are no longer available through direct download; instead, users are required to submit an access request.

\bibliography{Ref.bib}

\end{document}